\documentclass{ctr}

\usepackage{ctrfont}
\usepackage{natbib}

\usepackage{url}
\usepackage{amsmath}
\usepackage{amssymb}

\usepackage[dvips]{graphicx}
\usepackage{psfrag}
\usepackage{epsfig}
\usepackage{color}

\newcommand{\EE}[1]{\mathbb{E}\left[#1\right]}
\newcommand{\Var}[1]{\mathbb{V}ar\left(#1\right)}

\newcommand{\defin}{\stackrel{\mathrm{def}}{=}}

\renewcommand{\thefootnote}{\fnsymbol{footnote}}

\title{Multi-fidelity uncertainty quantification of irradiated particle-laden turbulence}
\shorttitle{Multi-fidelity uncertainty quantification}
\author{L. Jofre, G. Geraci\footnote{Sandia National Laboratories}, H.~R. Fairbanks\footnote{University of Colorado}, A. Doostan\thefootnote{} \and G. Iaccarino}
\shortauthor{Jofre et~al.}

\begin{document}


\maketitle

\section{Motivation and objectives}	\label{sec:motivation}

Uncertainty quantification (UQ) has become increasingly popular in the modeling and simulation community.
The ability to quantitatively characterize and reduce uncertainties, in conjunction with model verification and validation (V\&V), plays a fundamental role in increasing the reliability of numerical simulations.
In this regard, the Predictive Science Academic Alliance Program (PSAAP) II at Stanford University focuses on advancing the state of the art in large-scale predictive simulations of irradiated particle-laden turbulence relevant to concentrated solar power (CSP) systems.
To this end, physics-based models are developed and the model predictions are validated against data acquired from an in-house experimental apparatus designed to mimic a scaled-down particle-based solar energy receiver.

\subsection{Irradiated particle-laden turbulent flow}	\label{sec:irradiatedParticleLadenFlow}

Turbulent flow laden with inertial particles, or droplets, in the presence of thermal radiation is encountered in a wide range of natural phenomena and industrial applications.
For instance, it is well established that turbulence-driven particle inhomogeneity plays a fundamental role in determining the rate of droplet coalescence and evaporation in ocean sprays~\citep{Veron2015-A} and atmospheric clouds~\citep{Shaw2003-A}.
Another example is found when studying fires, in which turbulence, soot particles, and radiation are strongly interconnected resulting in very complex physical processes~\citep{Tieszen2001-A}.
From an industrial point of view, important applications include the atomization of liquid fuels in combustion chambers~\citep{Lasheras2000-A}, soot formation in rocket engines~\citep{Raman2016-A}, and more recently, volumetric particle-based solar receivers for energy harvesting~\citep{Ho2017-A}.

Even in the simplest configuration, e.g., homogeneous isotropic turbulence (HIT), particle-laden turbulent flow is known to exhibit complex interactions between the carrier and dispersed phases in the form of preferential concentration and turbulence modulation~\citep{Balachandar2010-A}.
Preferential concentration is the phenomenon by which heavy particles tend to avoid intense vorticity regions and accumulate in regions of high strain rate, while turbulence modulation refers to the alteration of fluid flow characteristics in the near-field region of particle clusters as a result of two-way coupling effects, e.g., enhanced dissipation, kinetic energy transfer, or formation of wakes and vortexes.
The physical complexity is further increased by the simple addition of solid walls as turbophoresis~\citep{Caporaloni1975-A}, i.e., tendency of particles to migrate towards regions of decreasing turbulence levels, becomes an important mechanism for augmenting the inhomogeneity in spatial distribution of the dispersed phase by accumulating particles at the walls.

As described above, characterization of particle-laden turbulent flow is a difficult problem by itself, with many experimental and numerical research studies devoted to this objective in the past decades~\citep{Squires1991-A,Wang1996-A,Sardina2012-A}.
However, the problem of interest in this work involves, in addition to particle-flow coupling, heat transfer from the particles to the fluid through radiation absorption.
The practical application motivating the study of this phenomena is the improvement of energy harvesting in volumetric particle-based solar receivers.
At present, most CSP technologies use surface-based collectors to convert the incident solar radiation into thermal energy.
In this type of systems, the energy is transferred to the working fluid downstream of the collection point via heat exchangers, typically resulting in large conversion losses at high temperatures.
By contrast, volumetric solar receivers continuously transfer the energy absorbed by particles directly to the operating fluid as they are convected through an environment exposed to thermal radiation.
This innovative technology is expected to increase the performance of CSP plants by avoiding the necessity of heat-exchanging stages, while requiring significantly high radiation-to-fluid energy transfer ratios.
This requirement imposes a very complex design constraint as the physical mechanisms governing irradiated particle-laden turbulent flow are still not fully comprehended.
In fact, to the best of the authors' knowledge, only few recent works~\citep{Zamansky2014-A,Frankel2016-A} can be found in which the physics of a transparent fluid laden with solid particles interacting with radiative heating is carefully analyzed.

\subsection{Objectives and organization of the work}	\label{sec:objectives}

The system studied in this work is based on a small-scale apparatus designed to reproduce the operating conditions of volumetric particle-based solar receivers.
As a consequence, many different uncertainties naturally arise, when trying to numerically investigate its performance in terms, for instance, of thermal output and efficiency.
Examples include incomplete characterization of particle-size distribution~\citep{Rahmani2015-A} and radiation properties~\citep{Frankel2017-A}, variability in radiation input and boundary conditions, and structural uncertainty inherent in the models utilized~\citep{Jofre2017-A}.
In addition to the large number of uncertainties, accurate predictions of the complex interaction of particle-laden turbulent flow with radiative heat transfer demands the utilization of expensive high-fidelity (HF) numerical simulations.
As an example, the cost of a medium-scale HF calculation of this problem requires approximately $500$k core-hours per sample on Mira supercomputer~\citep{Alcf2017-O}.
Therefore, if brute-force UQ techniques, e.g., Monte Carlo (MC) simulation with ${\cal O}(10^{3})$ samples, are to be performed, the total cost is of the order of $500$M core-hours, resulting in unaffordable UQ campaigns.
In this regard, the objective of this work is to investigate multi-fidelity UQ strategies on large-scale, multiphysics applications based on the PSAAP II solar receiver.

The paper is organized as follows.
In Sec.~\ref{sec:simulationMethod}, the physical models utilized to simulate irradiated particle-laden turbulent flow are described.
Then, in Sec.~\ref{sec:multifidelityStrategies}, the multi-fidelity strategies investigated are presented.
The UQ campaign is detailed next, Sec.~\ref{sec:campaign}, in terms of computational setup, uncertainties, and quantities of interest (QoIs) considered.
Afterwards in Sec.~\ref{sec:results}, the performance of the multi-fidelity estimators is analyzed.
Finally, the work is concluded and future directions are proposed in Sec.~\ref{sec:conclusions}.

\section{Modeling of irradiated particle-laden turbulent flow}	\label{sec:simulationMethod}

The PSAAP II overarching problem involves the interaction of particles and wall-bounded turbulent flow in a radiation environment.
The equations describing this type of flow are continuity, Navier-Stokes in the low-Mach-number limit, conservation of energy assuming ideal-gas behavior, Lagrangian particle transport, and radiative heat transfer.

\subsection{Variable-density turbulent flow}	\label{sec:flow}

The volumetric particle-based solar receiver operates at atmospheric pressure conditions in which air, the carrier fluid, is assumed to follow the ideal-gas equation of state (EoS), $P_{th}=\rho_{g} R_{air}T_{g}$, where $P_{th}$ is the thermodynamic pressure, $\rho_{g}$ is the density, $R_{air}$ is the specific gas constant for air, and $T_{g}$ is the temperature.
As indicated by the EoS, density varies with temperature.
However, the Mach number of the flow, $Ma=u/c$ with $u$ the local flow velocity and $c$ the speed of sound of the medium, for the range of velocities and temperatures considered is small ($Ma < 0.03$).
Therefore, the low-Mach-number approximation~ \citep{Esmaily2017-A} is utilized to separate the hydrodynamic part, $p \ll P_{th}$, from the total pressure, $P_{tot}=P_{th} + p$. This decomposition results in the following equations of fluid motion
\begin{align}
  & \frac{\partial \rho_{g}}{\partial t} + \nabla \cdot \left(\rho_{g} \textbf{u}_{g} \right) = 0,	\label{eq:continuity} \\
  & \frac{\partial \left( \rho_{g} \textbf{u}_{g} \right)}{\partial t} + \nabla \cdot \left( \rho_{g} \textbf{u}_{g} \otimes \textbf{u}_{g} \right) = -\nabla p + \nabla \cdot \mu_{g} \left[ \left( \nabla \textbf{u}_{g} + \nabla \textbf{u}^{\intercal}_{g} \right) - \frac{2}{3}(\nabla \cdot \textbf{u}_{g})\boldsymbol{I} \right] + \rho_{g}\textbf{g} + \textbf{f}_{TWC},	\label{eq:momentum} \\
  & \frac{\partial \left( \rho_{g} C_{v,g} T_{g} \right)}{\partial t} + \nabla \cdot \left( \rho_{g} C_{p,g} T_{g} \textbf{u}_{g} \right) = \nabla \cdot\left( \lambda_{g} \nabla T_{g} \right) + S_{TWC},	\label{eq:energy}  
\end{align}
where $\textbf{u}_{g}$ is the gas velocity, $\boldsymbol{I}$ is the identity matrix, $\textbf{g}$ is the gravitational acceleration, $\mu_{g}$ and $\lambda_{g}$ are the dynamic viscosity and thermal conductivity, $C_{v,g}$ and $C_{p,g}$ are the isochoric and isobaric specific heat capacities, and $\textbf{f}_{TWC}$ and $S_{TWC}$ are two-way coupling terms representing the effect of particles on the fluid and approximated as
\begin{equation}
  \textbf{f}_{TWC} = \sum_{p} m_{p}\frac{\textbf{u}_{p} - \textbf{v}_{p}}{\tau_{p}}\delta \left(\textbf{x} - \textbf{x}_{p} \right), \quad S_{TWC} = \sum_{p} \pi d_{p}^{2} h\left( T_{p} - T_{g} \right)\delta \left(\textbf{x} - \textbf{x}_{p} \right),	\label{eq:mometum_energy_TWC}
\end{equation}
where $m_{p} = \rho_{p}\pi d_{p}^{3}/6$ and $\textbf{v}_{p}$ are the particle mass and velocity, $\textbf{u}_{p}$ is the gas velocity at the particle location, $\tau_{p}=\rho_{p}d_{p}^{2}/(18 \mu_{g})$ is the particle relaxation time and $d_{p}$ the particle diameter, $\delta \left(\textbf{x} - \textbf{x}_{p} \right)$ is the Dirac delta function concentrated at the particle location $\textbf{x}_{p}$, $h = Nu \lambda_{g}/d_{p}$ is the gas-particle convection coefficient with $Nu$ the particle Nusselt number ---\thinspace the Biot number is $Bi = h d_{p}/\lambda_{p} \ll 1$ in this problem, and therefore particles are assumed to be isothermal \thinspace--- and $T_{p}$ is the particle temperature.

\subsection{Lagrangian particle transport}	\label{sec:particles}

The carrier fluid is transparent to the incident radiation. 
Hence, micron-sized nickel particles, i.e., the dispersed phase, are seeded into the gas to generate a non-transparent gas-particle mixture that absorbs and transfers, by means of thermal convection, the incident radiation from the particles to the gas phase.
The diameters of the particles are several orders of magnitude smaller than the smallest significant (Kolmogorov) turbulent scale, $\tau_{\eta}$, and the density ratio between particles and gas is $\rho_{p}/\rho_{g} \gg 1$.
As a result, particles are modeled following a Lagrangian point-particle approach with Stokes' drag as the most important force~\citep{Maxey1983-A}.
Their description in terms of position, velocity and temperature is given by

\begin{align}
  & \frac{d \textbf{x}_{p}}{d t} = \textbf{v}_{p},	\label{eq:position} \\
  & \frac{d \textbf{v}_{p}}{d t} = \frac{\textbf{u}_{p} - \textbf{v}_{p}}{\tau_{p}} + \textbf{g}, 	\label{eq:velocity} \\
  & \frac{d \left( m_{p} C_{v,p} T_{p} \right)}{d t} = \frac{\pi d_{p}^{2}\left(1 - \omega \right)}{4}\int_{4\pi}\left( I - \frac{\sigma T_{p}^{4}}{\pi} \right)d\Omega - \pi d_{p}^{2} h \left(T_{p} - T_{g} \right),	\label{eq:temperature}
\end{align}
where $C_{v,p}$ is the particle specific isochoric heat capacity, $\omega = Q_{s}/\left( Q_{a} + Q_{s} \right)$ is the scattering albedo with $Q_{a}$ and $Q_{s}$ the absorption and scattering efficiencies, respectively, $I$ is the radiation intensity, $\sigma$ is the Stefan-Boltzmann constant, and $d\Omega=\mbox{sin}\thinspace\theta d\theta d\phi$ is the differential solid angle.
In the conservation equation for particle temperature, Eq.~\ref{eq:temperature}, the first term on the right-hand-side accounts for the amount of radiation absorbed by a particle, while the second term represents the heat transferred to its surrounding fluid. 


\subsection{Radiative heat transfer}	\label{sec:radiation}

In the problem under consideration, the flow and particles time scales are orders of magnitude larger than the radiation time scale, which is related to the speed of light.
As a consequence, it can be assumed that the radiation field changes instantaneously with respect to temperature and particle distributions, i.e., radiation field is quasi-steady.
Under this assumption, and considering that air is transparent at all wavelengths and that absorption and scattering are determined solely by the presence of particles and solid boundaries, the radiative heat transfer equation becomes
\begin{equation}
  \hat{\textbf{s}}\cdot \nabla I = -\sigma_{e}I + \sigma_{a}\frac{\sigma T_{p}^{4}}{\pi} + \frac{\sigma_{s}}{4 \pi}\int_{4\pi}I \Phi d\Omega,	\label{eq:radiation}
\end{equation}
where $\hat{\textbf{s}}$ is the direction vector, $\sigma_{e} = \sigma_{a} + \sigma_{s}$ is the total extinction coefficient with $\sigma_{a}$ and $\sigma_{s}$ the absorption and scattering coefficients, respectively, and $\Phi$ is the scattering phase function that describes the directional distribution of scattered radiation.

The total extinction coefficient can be defined also in terms of absorption and scattering efficiencies as $\sigma_{e} = \left( Q_{a} + Q_{s} \right)\pi d_{p}^{2} n_{p} / 4$ with $n_{p}$ the local number density of particles.
Moreover, since it has been assumed gray radiation, $Q_{a} + Q_{s} \approx 1$ leading to $\omega \approx Q_{s}$, and as a result $\sigma_{a} \approx Q_{a}\pi d_{p}^{2} n_{p} / 4$ and $\sigma_{s} \approx Q_{s} \pi d_{p}^{2} n_{p} / 4$.

\section{Multi-fidelity accelerated sampling strategies}	\label{sec:multifidelityStrategies} 

In computational science and engineering, multiple physical/mathematical/numerical models with different features can be constructed to characterize a system of interest.
Typically, computationally expensive HF models are designed to describe the system with the degree of accuracy required by the problem under study, while low-fidelity (LF) models are formulated as less accurate, but relatively cheaper, representations.
Outer-loop problems, such as inference, UQ and optimization, require large numbers of model evaluations for different input values, resulting in unaffordable computational requirements in the case of large-scale, multiphysics calculations.
The objective of multi-fidelity methods, therefore, is to reduce the cost of the outer-loop problem by combining the accuracy of the HF models with the speedup achieved by the LF representations.
Different multi-fidelity UQ strategies exist in the literature; see for example the review by \cite{Peherstorfer2016-TR}.
However, due to the high-dimensional input space and the complexity of the conservation equations, this study is restricted to a reduced subset of acceleration strategies appertaining to surrogate-based MC-type sampling approaches.

As its name indicates, MC-type approaches are derived from the original Monte Carlo method, in which the expectation of the QoI $Q = Q(\mathbf{\xi})$, as a function of the stochastic input $\mathbf{\xi}$, is estimated via a sample average.
Let $\EE{Q}$ and $\Var{Q}$ denote the mean and variance of $Q$.
Given $N$ independent realizations of the stochastic input, denoted $\mathbf{\xi}^{(i)}$, the MC estimator of $\EE{Q}$ is defined as $\hat{Q}^{MC}_{N} \defin N^{-1} \sum_{i=1}^N Q^{(i)}$, where $Q^{(i)} \defin Q(\mathbf{\xi}^{(i)})$.
Although unbiased, the accuracy of $\hat{Q}_{N}^{MC}$, measured by its standard deviation $\sqrt{\Var{Q}/N}$, decays slowly as a function of $N$.
Therefore, for a fixed computational budget a viable alternative to increase the MC accuracy is to possibly replace $Q$ with other quantities with the same mean but reduced variances.

\begin{figure}
  \centerline{\includegraphics[width=0.85\textwidth]{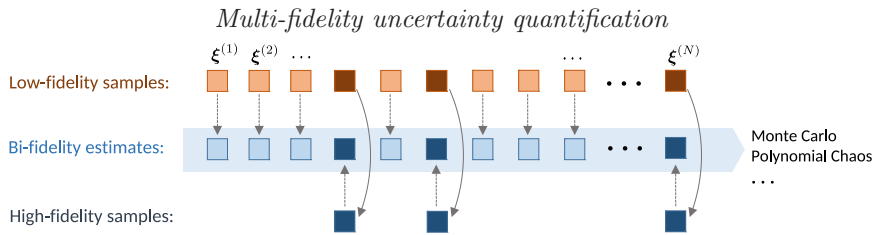}}
  \caption{Illustration describing the BF approximation method. First, a sweep of LF realizations (light-orange boxes) is performed from which the low-rank basis (dark-orange) and coefficient matrix are identified. Next, HF realizations of the LF-basis samples are computed (dark-blue). Finally, the BF approximation is obtained by applying the coefficient matrix to the HF basis (light-blue).} \label{fig:bfMethod}
\end{figure}

\subsection{Multi-level Monte Carlo}	\label{sec:multiLevel}

One of the most popular acceleration strategies is the multi-level (ML) method~\citep{Giles2008-A}.
This technique, inspired by the multigrid solver idea in linear algebra, is based on evaluating realizations of $Q$ from a hierarchy of models with different fidelity levels $\ell$, $\ell=0,\dots,L$ with $L$ the highest fidelity, in which $Q$ is replaced by the sum of differences $Y_\ell \defin Q_\ell - Q_{\ell-1}$, where by definition $Y_0 = Q_0$.
As a result, the QoI of the original and new ML problems have the same mean $\EE{Q}$.
An example of level is the grid resolution considered for solving the system of equations, so that a low- (or high-) fidelity model can be established by simulating $Q$ on a coarse (or fine) grid.
Then, $\EE{Q}$ can be computed using the ML QoI and an independent MC estimator on each level $\ell$ as
\begin{equation}
  \hat{Q}^{\mathrm{ML}} = \sum_{\ell=0}^{L} \hat{Y}^{\mathrm{MC}}_{\ell} = \sum_{\ell=0}^{L} \frac{1}{N_{\ell}} \sum_{i=1}^{N_{\ell}} Y^{(i)}_{\ell}.	\label{eq:mlEstimator}       
\end{equation}
This approach is referred to as Multilevel Monte Carlo (MLMC) and the resulting estimator has a variance equal to $\mathbb{V}ar\left(\hat{Q}^{\mathrm{ML}}\right)=\sum_{\ell=0}^{L} N_{\ell}^{-1} \mathbb{V}ar\left({Y_\ell}\right)$.
Consequently, if the level definition is such that $Q_\ell \rightarrow Q$ in mean square, then $\mathbb{V}ar\left({Y_\ell}\right) \rightarrow 0$ as $\ell \rightarrow \infty$.
Hence, fewer samples are required on the finer level $L$.
In particular, it is possible to show that the optimal sample allocation across levels, $N_\ell$, is obtained in closed form given a target variance of the MLMC estimator equal to $\varepsilon^2/2$ and resulting in
\begin{equation}
 N_\ell = \dfrac{ \sum_{k=0}^L \sqrt{ \mathcal{C}_k \mathbb{V}ar\left({Y_k}\right) } }{ \varepsilon^2/2 } \dfrac{ \mathbb{V}ar\left({Y_\ell}\right) }{ \mathcal{C}_\ell },
\end{equation}
where the cost of each evaluation $Y_\ell$ per level is denoted by $\mathcal{C}_\ell$. 

It is important to note that the variance decay can only be proven to be satisfied for levels based on a numerical discretization (spatial/temporal meshes) and not for general hierarchies of models, such as 1-D versus 2-D, Reynolds-averaged Navier-Stokes (RANS) versus large-eddy simulation (LES), etc.

\begin{figure}
  \centerline{\includegraphics[width=\textwidth]{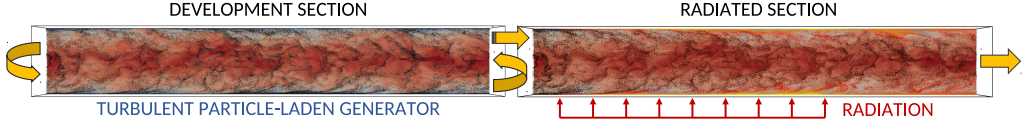}}
  \caption{Computational setup of the PSAAP II volumetric particle-based solar energy receiver. An isothermal periodic section (left domain) is utilized to generate fully developed particle-laden turbulent flow, which is used as inflow conditions for the second section (right domain) where the gas-particle mixture is irradiated perpendicularly to the flow direction from one the sides.}	\label{fig:problemSetup}
\end{figure}

\subsection{Multi-fidelity Monte Carlo}	\label{sec:multiFidelity}

To accommodate low-fidelity representations that are not obtained directly from coarsening the HF models, a common approach is to utilize LF realizations as a control variate~\citep{Pasupathy2014-A,Ng2014-A}.
In this work this strategy is referred to as multi-fidelity (MF).
In statistics, the control variate approach requires that a generic QoI $f$ is replaced by $f - \beta ( g - \EE{g} )$, where $g$ is a function chosen for its high correlation with $f$ and for which the value of $\EE{g}$ is readily available.
However, in the problem of interest here the low-fidelity model features are not available \textit{a priori}, and consequently need to be established during the computations along with the high-fidelity calculations.
As a consequence, the expected value of the low-fidelity model is generally approximated by means of a MC estimator requiring a set of additional (independent) LF computations.
The control variate MC estimator (multi-fidelity estimator) is defined as
\begin{equation}
  \hat{Q}^{\mathrm{MF}} = \hat{Q}^{\mathrm{HF,MC}} - \beta\left( \hat{Q}^{\mathrm{LF,MC}} - \EE{\hat{Q}^\mathrm{LF}} \right),	\label{eq:mfEstimator}                    
\end{equation} 
where the parameter $\beta$ is chosen to minimize the variance of $\hat{Q}^{\mathrm{MF}}$.
The optimal $\beta = \sqrt{\rho \left( \Var{Q^{\mathrm{HF}}} / \Var{Q^{\mathrm{LF}}} \right)}$ selection leads to 
\begin{equation}
 \Var{\hat{Q}^{\mathrm{MF}}} = \Var{ \hat{Q}^{\mathrm{HF,MC}}}  \left( 1 - \rho^{2} \dfrac{r}{1+r} \right),
\end{equation}
where $\rho^{2}$ is the correlation between the HF and the LF models, and $r$ is used to parametrize the additional $r N^{\mathrm{HF}}$ LF realizations needed in order to evaluate
\begin{equation}
 \EE{\hat{Q}^\mathrm{LF}} \approx \dfrac{1}{N^{\mathrm{HF}}(1+r)} \sum_{i=1}^{N^{\mathrm{HF}}(1+r)} Q^{\mathrm{LF},(i)}.
\end{equation}
As a result, the optimal control variate is obtained for a particular $r$ value which in turn depends on the correlation between the two models and their cost ratio.
In this report, the value is directly given by
\begin{equation}
r = -1 + \sqrt{ \dfrac{\mathcal{C}^\mathrm{HF}}{\mathcal{C}^{\mathrm{LF}}} \dfrac{\rho^2}{1-\rho^2}},	\label{eq:optimal_r}
\end{equation}
as described in~\cite{Geraci2015-A,Geraci2017-A}.
Moreover, a hybridization between the ML and MF approaches~\citep{Fairbanks2017-A} is also possible and will be studied in future works.

\begin{table}
\centering
\tabcolsep7pt\begin{tabular}{llll}
\hline
Variable & Interval & Variable & Interval \\
\hline
1. Prt. rest. coeff. 1 & [0.0 : 0.6] & 8. Mass load. ratio & [18 : 22]\%   \\
2. Prt. rest. coeff. 2 & [0.1 : 0.7] & 9. Prt. abs. eff.    & [0.37 : 0.41] \\
3. Prt. rest. coeff. 3 & [0.2 : 0.8] & 10. Prt. scatt. eff. & [0.69 : 0.76] \\
4. Prt. rest. coeff. 4 & [0.3 : 0.9] &  11. Radiation          & [1.8 : 2.0] $\mathrm{MW/m^{2}}$ \\
5. Prt. rest. coeff. 5 & [0.4 : 1.0] & 12. Radiated wall      & [1.6 : 6.4] $\mathrm{kW/m^{2}}$ \\
6. Stokes' drag corr.   & [1.0 : 1.5] &13. Opposite wall      & [1.2 : 4.7] $\mathrm{kW/m^{2}}$ \\
7. Prt. Nusselt num.   & [1.5 : 2.5] & 14. Side $x$-$y$ walls & [0.1 : 0.2] $\mathrm{kW/m^{2}}$ \\
\hline
\end{tabular}
\caption{List of random inputs with the corresponding ranges. All inputs are assumed to be uniformly distributed.}	\label{tab:uncertainties}
\end{table}

\subsection{Bi-fidelity low-rank approximation}	\label{sec:biFidelity}

An alternative methodology is the bi-fidelity (BF) approximation~\citep{Narayan2014-A,Doostan2016-A,Skinner2017-A,Hampton2017-A}, where a low-rank representation of the HF solution is generated using an ensemble of LF samples along with a relatively small number of selected HF samples.
By low-rank representation of a vector-valued QoI $\textbf{q}$, it is meant a linear approximation of $\textbf{q}$ in a small size basis $\{\textbf{q}(\xi^{(i)})\}$ consisting of selected realizations of $\textbf{q}$, i.e.,
\begin{equation}
\textbf{q}(\xi) \approx \sum_{i=1}^r \textbf{q}(\xi^{(i)}) c_i(\xi).
\end{equation}
Here, $c_i(\xi)$ are unknown coefficients, and the rank of approximation $r$ is assumed to be considerably smaller than the size of $\textbf{q}$.
The methodology to construct a BF approximation is illustrated in Fig.~\ref{fig:bfMethod}.
First, a sweep of $N$ MC simulations of the LF model is performed to generate the vector of LF realizations $\textbf{Q}^{\mathrm{LF}}=[ \textbf{q}^{\mathrm{LF}}_{\left(1 \right)} \dots \textbf{q}^{\mathrm{LF}}_{\left(N \right)}]$ (orange boxes).
The next step is to calculate its low-rank approximation, defined as $\textbf{Q}^{\mathrm{LF}}\approx[ \textbf{q}^{\mathrm{LF}}_{\left(i_{1} \right)} \dots \textbf{q}^{\mathrm{LF}}_{\left(i_{r} \right)}]\textbf{C}^{\mathrm{LF}}$ with $\textbf{q}^{\mathrm{LF}}_{i_{r}}$ the LF-basis vectors (dark-orange boxes) and $\textbf{C}^{\mathrm{LF}}$ the interpolation matrix.
This approximation can be obtained, for instance, by means of a rank-revealing QR algorithm, e.g.,~\citep{Gu1996-A,Cheng2005-A,Halko2011-A}.
Then, HF realizations with the same inputs as the LF-basis samples, $[\textbf{q}^{\mathrm{LF}}_{\left(i_{1} \right)} \dots \textbf{q}^{\mathrm{LF}}_{\left(i_{r} \right)}] \rightarrow [\textbf{q}^{\mathrm{HF}}_{\left(i_{1} \right)} \dots \textbf{q}^{\mathrm{HF}}_{\left(i_{r} \right)}]$, are computed (dark-blue boxes).
Finally, the BF approximation (light-blue boxes) is obtained by applying the interpolation matrix to the HF basis as
\begin{equation}
  \textbf{Q}^{\mathrm{BF}}=[ \textbf{q}^{\mathrm{HF}}_{\left(i_{1} \right)} \dots \textbf{q}^{\mathrm{HF}}_{\left(i_{r} \right)}]\textbf{C}^{\mathrm{LF}},	\label{eq:bfApproximation}
\end{equation}
and the estimates can then be used, for example, for MC estimation, or to form a polynomial chaos expansion (PCE)~\citep{Ghanem2003-B}.
For the analysis of this bi-fidelity construction, the interested reader is referred to~\cite{Hampton2017-A}.

\begin{table}
\centering
\tabcolsep7pt\begin{tabular}{cccccc}
\hline
Model & $\#$ HF runs & $\#$ LF runs & $\#$ Equivalent HF runs & \multicolumn{2}{c}{CI/mean [\%]} \\
\hline
HF    & 15      & -       & 15                 & \multicolumn{2}{c}{42.31}        \\
\hline
      &         &         &                    & ML    & MF                       \\
\hline
LF1   & 15      & 123     & 15.73              & 27.28 & 25.20                    \\
LF2   & 15      & 501     & 15.38              & 22.81 & 20.56                    \\
\hline
\end{tabular}
\caption{Computational cost and accuracy, defined as the ratio between confidence interval (CI) and mean, of the ML and MF estimators.}	\label{tab:mlmfPilot}
\end{table}

\section{Particle-based solar receiver uncertainty quantification campaign}	\label{sec:campaign}

\subsection{Computational setup and physical parameters}	\label{sec:setup_physics}

Numerical simulations of the volumetric particle-based solar receiver are performed on the computational setup depicted in Fig.~\ref{fig:problemSetup}.
Two square duct domains, with dimensions $1.7L\times W\times W$ ($L = 0.16$ m, $W = 0.04$ m) in the streamwise ($x$-axis) and wall-normal directions ($y$- and $z$-axis), are utilized to mimic the development and radiated sections of the experimental apparatus.
The development section (left domain) is an isothermal, $T_{0}=300$ K, periodic particle-laden turbulent flow generator that provides inlet conditions for the inflow-outflow radiated section (right domain).
The solid boundaries of the development section ($y$- and $z$-sides) are considered smooth, no-slip, adiabatic walls.
Regarding the radiated section, the same boundary conditions are imposed except for the radiated region in which the $y$- and $z$-boundaries are modeled as non-adiabatic walls accounting for heat fluxes due to the radiation energy absorbed by the glass windows.

The bulk Reynolds number of the gas phase at the development section is $Re_{b} = \rho_{g} u_{b} L/\mu_{g} = 20$k with $u_{b}$ the gas bulk velocity.
The particle-size distribution is approximated by 5 different classes with Kolmogorov Stokes numbers in the range $5<St_{\eta}=\tau_{p}/\tau_{\eta}<20$ and with a total mass loading ratio of $\mathrm{MLR} = n_{p}m_{p}/\rho_{g} \approx 20$\%.
The gas-particle mixture is volumetrically irradiated through a $L \times W$ glass window starting at $\Delta x=0.1 L$ from the beginning of the radiated section.
The radiation source consists of an array of diodes mounted on a vertical support placed $\Delta y = 2.875 W$ from the radiated window and aligned with the streamwise direction of the flow.
The diodes generate a total power of $P\approx 1$ kW approximately uniform within a $18^{\circ}$ cone angle.

\subsection{Uncertainties and quantities of interest}	\label{sec:uqs_qois}

The uncertainty quantification campaign considers $14$ stochastic variables to target experiment and model-form uncertainties, as shown in Table \ref{tab:uncertainties}.
These correspond to incertitude in particle restitution coefficient for the different classes ($1-5$), correction to Stokes' drag law ($6$), particle Nusselt number ($7$), mass loading ratio ($8$), particle absorption and scattering efficiencies ($9-10$), incident radiation flux ($11$), and heat fluxes from the walls to the fluid ($12-14$).

The intervals of the stochastic variables listed in Table~\ref{tab:uncertainties} have been carefully characterized based on information provided by the team responsible for conducting the experiments, and by taking into consideration results and conclusions extracted from published studies.
The intervals of the particle restitution coefficients follow the trend observed in experimental investigations by \cite{Yang2006-A} in which $C_{R}$ increases with Stokes number.
The expression for Stokes' drag force correction and its coefficient interval is based on the theoretical work by \cite{Brenner1962-A}.
The particle Nusselt number range is extracted from the numerical experiments of heated particles performed by~\cite{Ganguli2017-A}.
The intervals for particle absorption and scattering efficiencies are obtained from Mie scattering theory and take into account sensitivity to shape deformation as investigated by \cite{Farbar2017-A}.
Intervals for mass loading ratio, incident radiation, and heat fluxes from the walls to the fluid are characterized based on comparisons between preliminary numerical simulations and experimental results.

Time-averaged three-dimensional (3-D) solutions of the numerical simulations are saved in binary files from which first- and second-order statistics of different QoIs can be analyzed.
For example, gas velocity and density distributions, mean gas temperature and fluctuations, transmitted light, number density, velocity and temperature of particles, etc.
However, in this work, the performance of the multi-fidelity estimators is focused on thermal QoIs at a probe located $\Delta x = 0.3 L$ downstream from the radiated perimeter, and perpendicular to the flow direction along the $y$-axis at $z=W/2$.
Of particular interest, as these quantities are available from the experiments, are the time-averaged $y$-axis profile of gas temperature and gas heat flux over the streamwise-perpendicular plane, viz. radiation power transferred to the fluid.

\begin{table}
\centering
\tabcolsep7pt\begin{tabular}{cccc}
Model     & $\#$ HF runs & $\#$ LF runs & $\#$ Equivalent HF runs \\
\hline
LF2 (MF)  & 57      & 4064    & 61  \\
LF2 (ML)  & 68      & 6793    & 74  \\
\hline 
LF1 (MF)  & 81      & 1660    & 91  \\
LF1 (ML)  & 108     & 1364    & 117 \\
\hline 
HF        & 269     &         & 269 \\
\end{tabular}
\caption{Cost comparison -- 10\% accuracy in the prediction for mean $\Delta T$ ($32.03 \pm 1.6015$ K)}
\label{tab:mlmfPerformance}
\end{table}

\section{Performance analysis of multi-fidelity estimators}	\label{sec:results}

Three fidelity levels have been designed to perform the UQ campaign: one HF model and two LF representations, denoted LF1 and LF2.
The HF corresponds to a point-particle direct numerical simulation (PP-DNS) with sufficient resolution ($\sim55$M cells/ section) to capture all the significant (integral to Kolmogorov) turbulent scales, while approximating the particles as Lagrangian points ($\sim15$M particles/section) with nonzero mass.
The flow grid is uniform in the streamwise direction with spacings in wall units equal to $\Delta x^{+} \approx 12$, while stretched in the wall-normal directions with the first grid point at $y^{+}, z^{+} \approx 0.5$ and with resolutions in the range $0.5 < \Delta y^{+}, \Delta z^{+} < 6$.
The radiative heat transfer equation is solved on a discrete ordinates method (DOM) mesh of $270 \times 160 \times 160$ gridpoints ($\sim7$M cells) with $350$ quadrature points (discrete angles).

Based on the HF model, two LF models have been constructed by carefully coarsening the Eulerian and Lagrangian resolutions (allowing, in addition, for larger time steps), resulting in the LF1 and LF2 representations that are $\sim170\times$ and $\sim1300\times$ cheaper per sample than HF, respectively.
To explore the performance of the acceleration strategies on predicting temperature profile and heat flux at the probe location, a total of 16 HF, 128 LF1, and 512 LF2 samples derived from identical input realizations have been computed using supercomputing resources ($\sim 20$M core-hours) from Mira~\citep{Alcf2017-O}.

\begin{figure}
  \centering
  \includegraphics[width=0.475\textwidth]{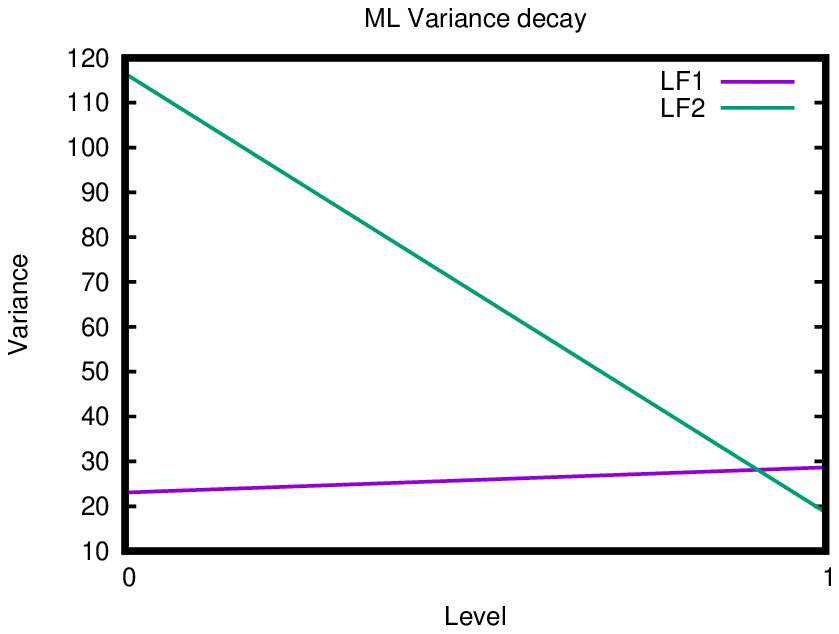}
  \includegraphics[width=0.475\textwidth]{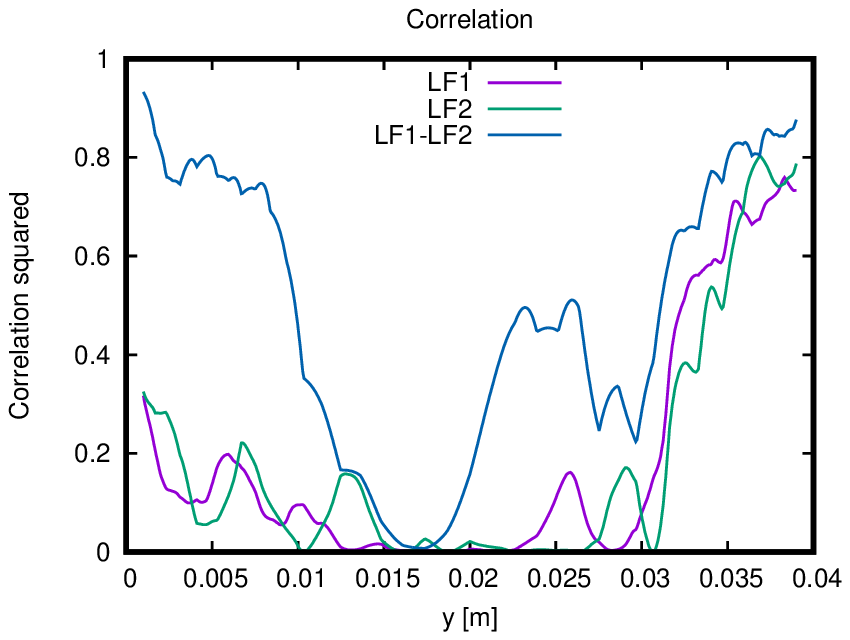}
\caption{ML variance decay across levels for time-averaged temperature at probe location $y \approx h$ (left). MF HF-LF correlation, $\rho^{2}$, for time-averaged probe temperature profile (right).}	\label{fig:mlmfEstimators}
\end{figure}

\subsection{Multi-level and multi-fidelity Monte Carlo estimators}	\label{sec:results_MCestimators}

A pilot study has been performed to understand the relations between the different models described above.
For this set of LF and HF realizations, the relevant variances and correlation for several QoIs have been computed.
Focus is placed here on the time-averaged temperature at probe location $y \approx h$, for which the variances and correlations are reported in Fig.~\ref{fig:mlmfEstimators}.

In the case of ML-LF1, the variance at the coarse level ($Y_{0}$) starts from a small value but it does not decay for the difference between LF1 and HF models ($Y_{1}$).
In contrast, the variance decays across levels for ML-LF2, however, starting from a much larger value at $Y_{0}$.
In terms of correlation with the HF, both LF1 and LF2 models present a similar behavior: correlation is small at the duct centerline, where the variability is almost negligible, 
but increases for the approximate range $0.75< y/h < 1$, where the variability of the HF estimator is larger.

The performance in terms of computational cost and estimator accuracy of the ML and MF strategies is summarized in Table~\ref{tab:mlmfPilot} for the time-averaged temperature at probe location $y \approx h$.
The general observation is that the utilization of the acceleration strategies is beneficial as the accuracy, defined as the ratio between confidence interval (CI) and mean, is reduced by half with practically negligible additional cost.
Particularly for this QoI, the best performance is obtained when using the MF-LF2 combination (bottom-right) due to the higher correlation shown in Fig.~\ref{fig:mlmfEstimators} and the reduced computational cost of the LF2 model.
In general, for this problem the LF2 model-based accelerations (either ML or MF) work consistently better than the ones based on LF1.
Based on these results, the performance of the strategies can be extrapolated to estimate what would be the cost of obtaining, for example, a temperature prediction at $y \approx h$ with a $10$\% CI.
The extrapolations, which are reported in Table~\ref{tab:mlmfPerformance} are based on the optimal allocation $N_\ell$ and additional LF ratio $r \approx 20, 70$ for LF1 and LF2, respectively.
For this particular case, $269$ HF calculations are needed, in contrast to only $57$ HF and $4064$ LF2 realizations, with an equivalent cost of $61$ HF runs, if the MF-LF2 strategy is utilized. Overall, due to a lack of monotonic variance decay, the MF methods outperform their ML counterparts.

\subsection{Bi-fidelity low-rank approximation performance}	\label{sec:results_BFestimator}

Two BF approximations are constructed to estimate the statistics for the QoI of time-averaged heat flux over the streamwise-perpendicular plane at the probe location. Bi-fidelity approximation 1 (BF1) is formed via samples from LF1 and HF models, and bi-fidelity approximation 2 (BF2) is formed via samples from LF2 and HF models. Both BF approximations are of rank $r=5$. The aim of this section is to compare how well the BF and LF models are able to capture the statistics of the HF heat flux QoI when limited samples are available.  

\begin{figure}
  \centerline{\includegraphics[width=\textwidth]{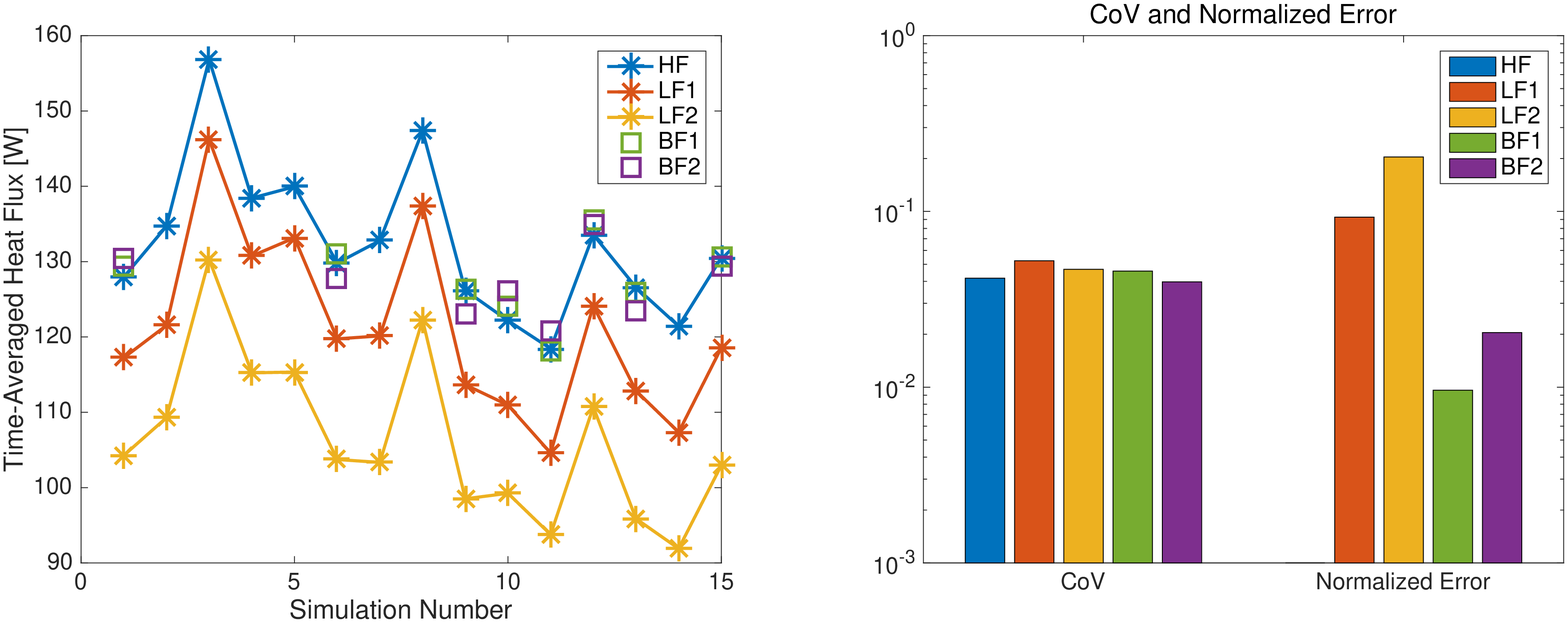}}
  \caption{Simulated heat flux values for all three fidelities and BF approximations (left). The BF1 and BF2 data displayed corresponds to non-basis data. CoV and normalized error values for the planar heat flux (right). Data based on 8 samples corresponding to available non-basis high-fidelity data.} \label{fig:bf_errors}
\end{figure}

Simulated values of the heat flux for all five models are displayed in Fig.~\ref{fig:bf_errors}(left). For the BF models, only the non-basis data is provided. Note that the LF1 and LF2 simulations are not able to accurately capture the correct heat flux values of the HF data; however, the LF1 and LF2 data are well correlated with the HF data. Due to this high correlation, the resulting BF1 and BF2 samples closely align with those of the HF samples, with BF1 data being slightly more accurate than BF2. 

Fig.~\ref{fig:bf_errors}(right) provides the normalized validation error between the LF and BF data sets and the HF data. The normalized error is defined to be the $\ell_2$ error of the scalar heat flux values relative to the $\ell_2$ norm of the HF heat flux values. The coefficient of variation (CoV) for the HF, LF, and BF data sets present values of same order, which is necessary to form an accurate approximation. The normalized validation errors show that the BF approximations are $10\times$ more accurate than their respective LF approximations, with errors of $1\%$ for the BF1 data and $2\%$ for the BF2 data. This is an indication that, based on available data, the BF models provide a more accurate representation of the HF data than the corresponding LF models. It is also of value to note that, while the LF2 model has a small error, the BF2 data is still an improvement over LF1. The relationship between LF1 and LF2, where LF1 is better than LF2, is consistent for their corresponding BF approximations as well. 

\begin{figure}
  \centerline{\includegraphics[width=1.1\textwidth]{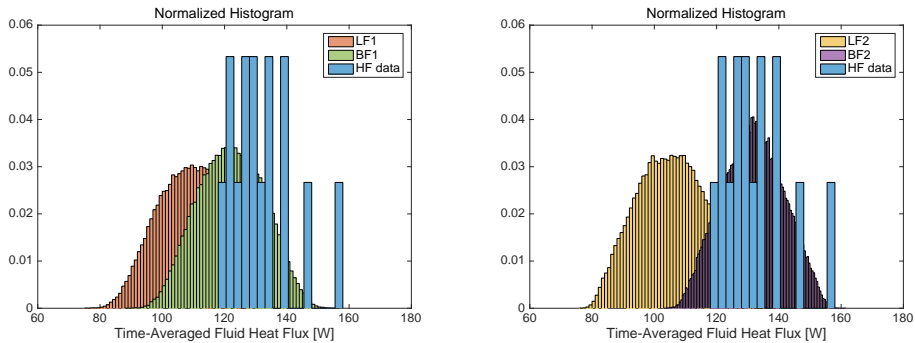}}
  \caption{Normalized histogram of mean heat flux at probe location based on sparse PCEs of (left) LF1 and BF1, and (right) LF2 and BF2. Using all available LF and BF data, 1000 surrogate values evaluated to construct the histograms.} \label{fig:bf_histogram}
\end{figure}

To further compare the LF and BF models, a first order PCE is computed from the available samples and used to generate additional samples. With this PCE surrogate, $1000$ samples are generated to construct a histogram. Fig.~\ref{fig:bf_histogram}(left) displays the normalized histograms of the LF1 and BF1 PCE surrogates along with the HF data and Fig.~\ref{fig:bf_histogram}(right) displays the normalized histograms of the LF2 and BF2 PCE surrogates along with the HF data. From both figures, it is observed that the BF generated histogram is more aligned with the HF data than the LF counterparts.
While previous results indicate that the BF1 approximation is a more accurate estimate of the HF data, the spread of the BF2 histogram appears to capture the spread of the HF data better than the BF1 histogram.

\section{Conclusions}	\label{sec:conclusions}

Performing UQ studies of large-scale, multiphysics applications is challenging due to the expensive HF calculations required and the large number of uncertainties encountered.
For instance, extrapolation of the PSAAP II UQ campaign to the full-system scale could cost on the order of one billion core-hours in some of the most advanced supercomputers.
Therefore, ML, MF and BF strategies have been explored in this work to effectively reduce the cost of such studies.

Based on the system of interest and methods considered, the MF performs better than the ML due to the high correlation but significant bias of the LF models.
In terms of the BF approach, the two approximations similarly outperform, by an order of magnitude, their associated LF estimators despite the larger error of the LF2 data, which is an indicator of the robustness of the methodology.
An interesting strategy would be the hybridization of these two approaches by utilizing the BF methodology as a control variate. 

Ongoing and future work focuses on improving the performance of these strategies for Lagrangian particles in the context of more challenging mass loading ratio and radiation inputs.
In parallel, data compression strategies for particle-laden flow are being explored, as well as better understanding of physical versus stochastic variability convergence.

\section*{Acknowledgments} 

This investigation was funded by the Advanced Simulation and Computing (ASC) program of the US Department of Energy's National Nuclear Security Administration via the PSAAP II Center at Stanford under contract DE-NA-0002373.

Sandia National Laboratories is a multimission laboratory managed and operated by National Technology and Engineering Solutions of Sandia, LLC., a wholly owned subsidiary of Honeywell International, Inc., for the U.S. Department of Energy's National Nuclear Security Administration under contract DE-NA-0003525.

This material is based upon work of Prof. A. Doostan supported by the U.S. Department of Energy Office of Science, Advanced Scientific Computing Research, under award DE-SC0006402, and NSF grant CMMI-145460. 

An award of computer time was provided by the ASCR Leadership Computing Challenge (ALCC) program.
This research used resources of the Argonne Leadership Computing Facility (ALCF), which is a DoE Office of Science User Facility supported under contract DE-AC02-06CH11357.
This research also used resources of the Oak Ridge Leadership Computing Facility (OLCF), which is a DoE Office of Science User Facility supported under contract DE-AC05-00OR22725.

The authors would like to acknowledge J.~H. Kim, A.~J. Banko, Dr. L. Villafa\~ne and Prof. J.~K. Eaton for providing the experimental data of the Stanford University PSAAP II volumetric particle-based solar receiver, Dr. T.~D. Economon for the help provided with the computations, and S. Ganguli, J. Horwitz, Dr. A. Frankel, Dr. M. Esmaily, and Dr. J. Urzay for the fruitful discussions regarding the uncertainty intervals. 

\bibliographystyle{ctr}


\end{document}